\documentclass[a4paper,11pt]{article}
\usepackage{amsmath}

\def\bgfrac#1#2{\displaystyle \frac{\displaystyle {#1}}{\displaystyle {#2}}}

\def\tfrac#1#2{\displaystyle{\frac{\partial #1}{\partial #2}}}
\def\ud{\text{d}}

\def\rs{\ensuremath{r^*} }
\title{\LARGE Motion of Massive and Massless Test particles\\ in Dyadosphere Geometry}

\author{B. Raychaudhuri\thanks{Surya Sen Mahavidyalaya, Siliguri, India. E-mail: biplab.raychaudhuri@gmail.com}, F. Rahaman\thanks{Department of Mathematics, Jadavpur University, Kolkata, India, E-mail: farook-rahaman@yahoo.com}, M. Kalam\thanks{Netaji Nagar College for Girls, Kolkata, India, E-mail: mehedikalam@yahoo.co.in} and A. Ghosh\thanks{Department of Mathematics, Jadavpur University, Kolkata, India}}

\begin{document}

\date{}

\maketitle

\begin{abstract}
Motion of massive and massless test particle in equilibrium and non-equilibrium case is discussed in a dyadosphere geometry through Hamilton-Jacobi method. Geodesics of particles are discussed through Lagrangian method too. Scalar wave equation for massless particle is analyzed to show the absence of superradiance.
\end{abstract}


\section{Introduction}
It is now well established that a black hole is uniquely characterized by its mass $M$, charge $Q$ and angular momentum $J$. The exterior solution for a rotating charged black hole is given the Kerr-Newman metric while that for the uncharged one is given by Kerr geometry. Further, if the black hole is charged but non-rotating, then the solution is Reissner-Nordstr\"om metric, while for uncharged case the black hole is represented by the Schwarzschild geometry. 

Recently Ruffini~\cite{ruffini} and Preparata, Ruffini and Xue~\cite{preparata} discussed a special region just outside the horizon of charged black holes where the electric field goes beyond its classical limit. This is a situation where effects of vacuum fluctuations should be considered. When the electromagnetic field exceeds the Heisenber-Euler~\cite{HE} critical value for $e^+ e^-$ pair production, in a very short time of the order $\sim \mathcal{O}(\hbar/mc^2)$, a very large number of $e^+e^-$ pair is created there. This reaches in thermodynamic equilibrium with a photon gas. The period of overcritical state is of approximately $10^7$~s.~\cite{preparata}. The dissipation occurs through a process involving the \textit{dyadosphere} of a black hole.

 Taking into account the one-loop correction due to QED to the first order of approximation the space time geometry has the form
\begin{equation}\label{metric}
 \ud s^2 = -f(r) \ud t^2 + \bgfrac{\ud r^2}{f(r)} + r^2\left( \ud \theta + \sin^2 \theta \ud \phi^2\right),
\end{equation}
where
\begin{equation}\label{fr}
 f(r) = 1 - \frac{2M}{r} + \frac{Q^2}{r^2} - \frac{\sigma Q^4}{5r^6}.
\end{equation}
$\sigma$ is a coupling constant. The metric reduces to usual charged non-rotating RN metric in the limit $\sigma=0$.  

Preparata, Ruffini and Xue~\cite{preparata} discussed some inequalities which must be satisfied during the forming process of a dyadosphere of a black hole. The authors put forward a theoretical model on the concept of dyadosphere explaining Gamma-Ray-Bursts phenomenon. Lorenci et. al.~\cite{lorenci} discussed the expression for the bending of light in the dyadosphere of the charged black hole.

In this paper we study the motion of test particle in the field of the dyadosphere of a black hole. The procedure outlined by Rahaman~\cite{farook} is followed here. In Sec.~\ref{HJ} we discuss both static and nonstatic cases for charged and uncharged test particle~\cite{farook}. Sec.~\ref{teststatic} will discuss the motion of test particles in static equilibrium while Sec.~\ref{testnonequi} will describe the motion of test particles in non equilibrium state. We discuss the wave  equation for a massless scalar wave in Sec.~\ref{waveeqn} to split and solve  the radial and angular equations. Absence of superradiance is expectedly observed in this case. Finally in Sec.~\ref{lagrangian}   geodesics of dyadosphere metric using Lagrangian technique is discussed.

\section{Trajectory of Test Particle through Hamilton- Jacobi Formalism}\label{HJ}

We first consider a test particle with mass $m$ and charge $e$ moving in the modified gravitational field of a RN black hole due to the presence of a dyadosphere (henceforth called \textit{dyadosphere field}). The Hamilton Jacobi equation for the test particle is given by~\cite{landau}
\begin{equation}\label{eqHJ}
 g^{ik} \left( \tfrac{S}{x^i} + e A_i \right) \left( \tfrac{S}{x^k} + e A_k \right) +m^2 =0,
\end{equation}
where $g_{ik}$  is the  background geometry, i.e. the metric~\eqref{metric} and 
\begin{equation}
 A_i = \frac{Q \ud t}{r} \longrightarrow \text{gauge potential.} 
\end{equation}

$S$ is the  usual Hamilton's characteristic function.

For the metric~\eqref{metric} the explicit form of the Hamilton-Jacobi equation is given by~\cite{farook}
\begin{equation}
-\frac{1}{f} \left( \tfrac{S}{t} + \frac{e Q}{r} \right)^2 + f \left( \tfrac{S}{r} \right)^2 + \frac{1}{r^2}\left( \tfrac{S}{\theta}\right)^2 + \frac{1}{r^2 \sin^2 \theta} \left( \tfrac{S}{Q} \right)^2 + m^2 =0. 
\end{equation}

 To solve this partial differential equation, an ansatz is proposed~\cite{farook}. 

\begin{equation}\label{ansatz}
 S(t,r,\theta,\phi) = -E.t + S_1(r) + S_2(\theta) + \mathbf{J. \phi}. 
\end{equation}

This will enable us to separate the variables.  $E$ and $J$ are identified as the energy and angular momentum of the particle. The substitution of this ansatz in the previous equation yields

\begin{equation}\label{sej}
 \begin{array}{lll}
\tfrac{S}{t} = -E, &  & \tfrac{S}{r} = \tfrac{S_1}{r}, \\[.3cm]
\tfrac{S}{\theta} = \tfrac{S_2}{\theta}, &  & \tfrac{S}{Q} = J.
\end{array}
\end{equation}

Substituting we obtain the following equations

\begin{equation}
 -\frac{1}{f} \left( -E+ \frac{eQ}{r} \right)^2 + \left( \tfrac{S_1}{r}\right)^2 + \frac{1}{r^2} \left(\tfrac{S_2}{\theta} \right)^2 + \frac{1}{r^2\sin^2 \theta} J^2 +m^2 =0, 
\end{equation}
and
\begin{equation}
 -\frac{r^2}{f}\left( E - \frac{eQ}{r} \right)^2 + fr^2 \left( \tfrac{S_1}{r} \right)^2 + \left( \tfrac{S_2}{\theta}\right)^2 + m^2r^2 +\frac{1}{\sin^2\theta} J^2 =0.
 \end{equation}

Now,
\begin{equation}
 -\frac{r^2}{f}\left( E - \frac{eQ}{r} \right)^2 + fr^2 \left( \tfrac{S_1}{r} \right)^2 +m^2r^2 = -p^2,
\end{equation}
giving
 \begin{equation}
 \left( \tfrac{S_1}{r} \right)^2 = \frac{\left(E - \frac{eQ}{r}\right)^2}{f^2} - \frac{m^2}{f} - \frac{p^2}{r^2f}.
\end{equation}

With all these, we find the following solutions for the functions $S_1(r)$ and $S_2(\theta)$ as
\begin{equation}
 S_1(r) = \epsilon \int{\left[ \frac{\left(E - \frac{eQ}{r} \right)^2}{f^2} - \frac{m^2}{f} \frac{p^2}{fr^2} \right]^{1/2} \ud r },
\end{equation}
and 
\begin{equation}
 S_2(\theta) = \epsilon \int \left[ p^2 - J^2 \csc^2\theta \right]^{1/2} \ud\theta,
\end{equation}
where $\epsilon =\pm 1$.

To determine the trajectory of the particle through Hamiltonian Jacobi method, we consider~\cite{farook}
\begin{equation}
 \tfrac{S}{E} = \text{constant}, \qquad \tfrac{S}{J} = \text{constant}, \qquad \tfrac{S}{P} = \text{constant}.
\end{equation}

Without any loss of generality the constants may be considered to be zero. Then, from Eq.~\eqref{ansatz} we obtain
\begin{equation}
\tfrac{S}{E} = -t
\end{equation}
which gives
\begin{equation}\label{tsoln}
 t= \epsilon \int{ \frac{1}{f^2} \left(E - \frac{eQ}{r}\right) \left[ \frac{1}{f^2}\left(E-\frac{eQ}{r}\right)^2 -\frac{m^2}{f} - \frac{p^2}{fr^2} \right]^{-1/2} \ud r},
\end{equation}

\begin{equation}\label{phisoln}
 \phi = \epsilon \int \left( J \csc^2\theta \right) \left[ p^2 - J^2 \csc^2 \theta\right]^{1/2} \ud \theta,
\end{equation}
and,
\begin{equation}
 \cos^{-1}\left( \frac{\cos \theta}{\nu}\right) = \int \frac{1}{fr^2} \left[ \frac{1}{f^2} \left(E-\frac{eQ}{r^2}\right)^2 -\frac{m^2}{f} - \frac{p^2}{fr^2}\right]^{1/2} \ud r,
\end{equation}
where $\nu^2 = 1- \bgfrac{J^2}{p^2}$.

The Radial velocity is particle is obtained from Eq.~\eqref{tsoln}
\begin{equation}\label{radvel}
 \frac{\ud r}{\ud t} = f^2\left(E - \frac{eQ}{r^2}\right)^{-1} \left[ \frac{1}{f^2} \left(E-\frac{eQ}{r} \right)^2 - \frac{m^2}{f} - \frac{p^2}{fr^2} \right].
\end{equation}

The turning point of the particle is $\bgfrac{\ud r}{\ud t} =0$ and the potential curves are
\begin{equation}
 \left(E-\frac{eQ}{r} \right)^2 - m^2f - \frac{p^2}{r^2}f =0. 
\end{equation}

\begin{figure}[htbp]
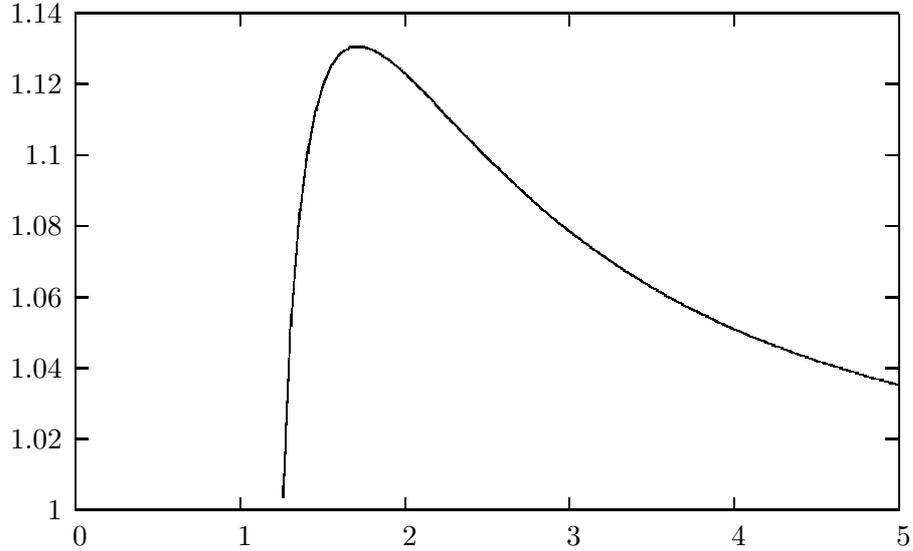

\begin{center}
\input veff1.tex
\end{center}
\caption{$V(r)$ for $e/m = 1$, $Q=M=1$, $\sigma=0.5$ and $p/m = 1.5$}\label{figveff1}
\end{figure}

Solving 
\[E= \frac{eQ}{r} + \sqrt{f} \left( m^2 + \frac{p^2}{r^2} \right)^{1/2} \]
The effective potential is given by
\[ V(r) = \frac{E}{m} = \frac{eQ}{mr} + \sqrt{f(r)} \left( 1+\frac{p^2}{m^2r^2} \right)^{1/2} \]
as an explicitly function of $r$ only. A representative curve is given below (Fig.~\ref{figveff1}). We have considered the specific charge of the particle $e/m=1$, and $p/m=1.5$ and $\sigma=0.5$ for convenience. This is a plot for a extremal case where $Q=M$. We set them to unity.

\begin{figure}[htbp]
 \begin{center}
  \input{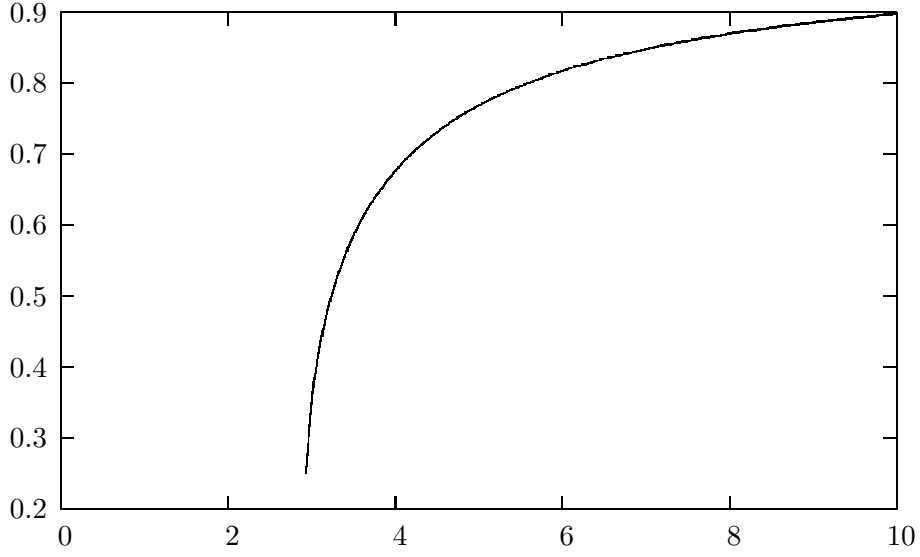}
 \end{center}
\caption{$V(r)$ for $e/m = 1$, $Q=0.5$, $M=1.5$, $\sigma=0.5$ and $p/m = 1.5$}\label{figveff2}
\end{figure}

Different curves may be drawn depending upon different values of the parameter.
We give another plot where $M=1.5$ but $Q=0.5$ (Fig.~\ref{figveff2}).

In a stationary system, $E$ and so $V(r)$ must have an extremal value. Hence the value of $r$ for which the energy attains its extremal value is given by the solution of
\[\frac{\ud V}{\ud r} =0. \]
This gives
\begin{equation}
- \frac{eQ}{mr^2} + \frac{1}{2\sqrt{f}}\left(1 + \frac{p^2}{m^2r^2}\right)^{1/2}\tfrac{f}{r} -  \sqrt{f} \left(1+\frac{p^2}{m^2r^2}\right)^{-1/2} \frac{p^2}{m^2r^3} =0.
\end{equation}

We obtain
\[
\frac{eQ}{mr^2}f^{1/2}\left( 1+ \frac{p^2}{m^2r^2} \right)^{1/2} = \frac{1}{2} \left(1 + \frac{p^2}{m^2r^2} \right) \tfrac{f}{r} - f\frac{p^2}{m^2r^3}.\]

With the expression of $f(r)$ the equation becomes
\begin{equation}\label{eqgeneral}
\begin{split}
\frac{eQ}{m}\left( 1-\frac{2M}{r}+\right. & \left. \frac{Q^2}{r^2} - \frac{\sigma Q^4}{5 r^6}\right)^{1/2} \left( 1 + \frac{p^2}{m^2r^2} \right)^{1/2} =  \\
 &\left(M - \frac{Q^2}{r} + \frac{3}{5}\frac{\sigma Q^4}{r^5} \right) \left(1 + \frac{p^2}{m^2r^2} \right) - \frac{p^2}{m^2 r} \left( 1 - \frac{2M}{r} + \frac{Q^2}{r^2} - \frac{\sigma Q^4}{5r^6} \right).
 \end{split}
\end{equation}
This expression is used in further discussions below.


\section{Test particle in Static Equilibrium}\label{teststatic}
In static equilibrium, momentum $p$ in Eq.~\eqref{eqgeneral} must be zero. So, the value of $r$ for which potential will be an extremal is given by
\[
\frac{eQ}{m} \left( 1 - \frac{2M}{r} + \frac{Q^2}{r^2} - \frac{\sigma Q^4}{5r^6} \right)^{1/2} = M - \frac{Q^2}{r}+ \frac{3}{5}\frac{\sigma Q^4}{r^5} 
\]
Simplifying and arranging 
\begin{equation}
\begin{split}
\left(M^2 - \frac{e^2Q^2}{m^2} \right) r^{10}& - 2MQ^2\left(1-\frac{e^2}{m^2}\right)r^9+ \left(1-\frac{e^2}{m^2}\right)Q^4r^8 \\
&+\frac{6}{5} M\sigma Q^4r^5 - \frac{\sigma Q^6}{5} \left(1-\frac{e^2}{m^2}\right) r^4 + \frac{9}{25} \sigma^2 Q^8 =0
\end{split} 
\end{equation}

This is a 10th order equation in $r$. If $\bgfrac{e^2}{m^2}>\bgfrac{M^2}{Q^2}$, then the last term is negative, i.e at least one positive root exists. If on the other hand, $\bgfrac{e^2}{m^2}=\bgfrac{M^2}{Q^2}$, then the above equation reduces to a ninth degree equation with negative last term implying that a real positive root exists. In other words, the particle will be trapped by the dyadosphere field.

\begin{figure}[htbp]
 \begin{center}
  \input{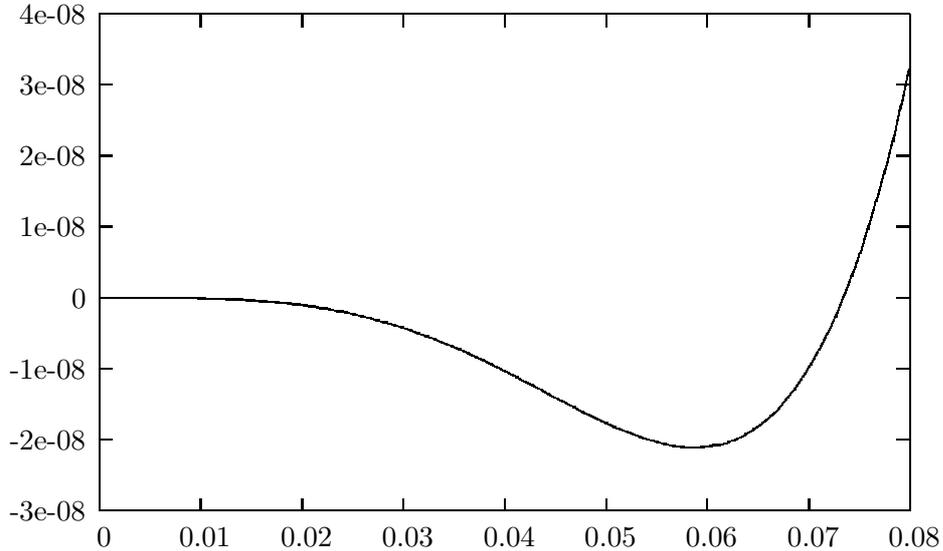}
 \end{center}
\caption{$M=1$, $e/m=0.5$, $Q=0.7$ and $\sigma = 0.5$}\label{figr10}
\end{figure}

Note that if the particle is of unit specific charge, $e/m=1$ then the equation becomes

\begin{equation}
 \left(M^2 - Q^2\right) r^{10} + \frac{6}{5×}M\sigma Q^4 r^5 + \frac{9}{25×}\sigma^2Q^8=0,
\end{equation}
i.e. for an extremal RN field, the equation becomes a 5th degree equation on $r$.
The function is plotted below (Fig.~\ref{figr10}). Again we demonstrate the nature of the function with $M=1$, $e/m=0.5$, $Q=0.7$ and $\sigma=0.5$. The function is monotonically increasing outside the range specified.

\section{Test particle in Non equilibrium state, $p\neq0$}\label{testnonequi}

\subsection{Case I -- Uncharged test particle $(e=0)$}
We first discuss the case of uncharged test particle $e=0$. In the nonequilibrium case, $p\neq 0$. The expression~\eqref{eqgeneral} becomes
\begin{equation}\label{noneqe0}
 \left(M -\frac{Q^2}{r} + \frac{3}{5}\frac{\sigma Q^4}{r^5}\right)\left(1+\frac{p^2}{m^2r^2}\right)=\frac{p^2}{m^2r} \left( 1 -\frac{2M}{r} + \frac{Q^2}{r^2} - \frac{\sigma Q^4}{5 r^6}\right)
\end{equation}

Simplifying and rearranging we obtain an algebraic equation of degree 7.
\begin{equation}
Mr^7 - \left( Q^2 + \frac{p^2}{m^2} \right) r^6 + 3 \frac{p^2}{m^2}M r^5 -   2\frac{p^2}{m62} Q^2r^4  
  + \frac{3}{5} \sigma Q^4 r^2 + \frac{4}{5} \sigma Q^4 \frac{p^2}{m^2} =0
\end{equation}

Since there are four  changes of sign in the above equation, then by Descartes rule of signs, the equation can have 4 positive roots.

\subsection{Case II, Charged particle $(e \neq 0)$}
In case of charged particle $e\neq 0$ and the Eq.~\eqref{eqgeneral} becomes
\begin{equation}
 \begin{split}
\left( M - \frac{Q^2}{r} + \frac{3}{5} \frac{\sigma Q^4}{r^5} \right)& \left(1 + \frac{p^2}{m^2r^2} \right) - \frac{p^2}{m^2 r} \left( 1 - \frac{2M}{r} + \frac{Q^2}{r^2} - \frac{\sigma Q^4}{5 r^6} \right)\\
& - \frac{eQ}{m} \left( 1 -\frac{2M}{r} + \frac{Q^2}{r^2} - \frac{\sigma Q^4}{5 r^6} \right)^{1/2} \left( 1 + \frac{p^2}{m^2r^2} \right)^{1/2}=0.
\end{split}
\end{equation}

This is an equation of degree 14 in $r$. When explicitly written this expression contains 60 terms. In terms of descending power of $r$ it reduces to

\begin{equation}
\begin{split}
 \left( M^2 - \frac{e^2Q^2}{m^2}\right)& r^{14}  - 2M \left(Q^2 + \frac{p^2}{m^2} - \frac{e^2Q^2}{m^2} \right)r^{13}  \\
& +\frac{p^2}{m^2}\left(2M^2+4M+\frac{m^2}{p^2}Q^4 -2Q^2 +\frac{p^2}{m^2} -\frac{e^2Q^2}{m^2} \right) r^{12} \\
& \ldots + 0 \times r + \frac{8}{5}\frac{p^4}{m^4} \sigma^2 Q^8 =0
\end{split}
\end{equation}

In this case too, if the specific charge of the test particle $e/m$ is greater than $M/Q$ then the last term is negative, i.e. at least one positive root exists. If on the other hand $\bgfrac{e}{m} = \bgfrac{M}{Q}$, then the above equation reduces to a 13 degree equation. with negative last term if $M^2 - Q^2 - \bgfrac{p^2}{m^2} >0$ implying the existence of  a real positive root. In these cases the particle will be trapped by the dyadosphere. 


\section{Solution of massless Scalar Wave Equation in Dyadosphere metric}\label{waveeqn}
In this section we shall analysis the scalar wave equation for dyadosphere geometry following Brill et. al~\cite{brill} who first showed that variables can be separated for Kerr geometry. (We consider massless scalar field here; extension to massive field is straightforward and devoid of any further conclusion.) 
The wave equation for a massless particle is given by
\begin{equation}
 g^{-1/2} \tfrac{}{x^\mu} \left( g^{1/2} g^{\mu\nu} \tfrac{}{x^\nu} \right) \chi  =0
\end{equation}

Here $g_\mu\nu$ is given by Eq.~\eqref{metric}. If we now put all the values we obtain the following equation
\begin{equation}
 -\frac{r^4 \sin\theta}{\Delta}\frac{\partial^2 \chi}{\partial t^2} + \sin \theta \tfrac{}{r}\left(\Delta \tfrac{\chi}{r}\right) + \tfrac{}{\theta}\left( \sin\theta \tfrac{}{\theta} \right)\chi + \frac{1}{\sin\theta}\frac{\partial^2 \chi}{\partial \phi^2}=0
\end{equation}

Where $\Delta = r^2 -2Mr+Q^2 - \bgfrac{\sigma Q^4}{5 r^4}$.

This equation be solved by separation of variable with the ansatz
\begin{equation}\label{chi}
 \chi = e^{-i\omega t} e^{im\phi} R(r) \Theta(\theta)
\end{equation}

Putting this in the wave equation we obtain
\[
\frac{r^4 \sin \theta}{\Delta}\omega^2 \chi +  \frac{\sin \theta}{R}  \tfrac{}{r}\left(\Delta \tfrac{R}{r}\right) \chi
+ \frac{1}{\Theta} \tfrac{}{\theta}\left(\sin\theta\tfrac{\Theta}{\theta} \right) \chi- \frac{m^2}{\sin\theta} \chi =0 
\]

Thus the variables can be easily separated. The radial equation reduces to

\begin{equation}\label{radial}
\Delta \tfrac{}{r}\left( \Delta \tfrac{R}{r} \right) + (r^4 \omega^2 - \Delta \lambda)R=0 
\end{equation}

The angular part becomes
\begin{equation}
 \frac{1}{\sin\theta}\tfrac{}{\theta} \left(\sin\theta\tfrac{\Theta}{\theta}\right) - \frac{m^2}{\sin^2\theta}\Theta + \lambda \Theta =0
\end{equation}

With a substitution $x=\cos\theta$, the equation becomes

\begin{equation}
 (1-x^2)\frac{\ud^2 \Theta}{\ud x^2} - 2x \frac{\ud \Theta}{\ud x} - \left( \frac{m^2}{1-x^2} - \lambda \right) \Theta=0
\end{equation}

If we now write $\lambda = l(l+1)$ and demand that $l$ takes only integer values then the equation 
\begin{equation}\label{angular}
 (1-x^2)\frac{\ud^2 \Theta}{\ud x^2} - 2x \frac{\ud \Theta}{\ud x} + \left(  l(l+1) - \frac{m^2}{1-x^2}  \right) \Theta=0
\end{equation}
is Associated Legendre equation and the solution is give by the associated Legendre polynomial $ P^m_l(x)$ and is expressed as
\begin{equation}
\Theta^m_l(\cos\theta) = P^m_l(x) = \frac{(-1)^m}{2^l l!}\left( 1-x^2 \right)^{m/2} \frac{\ud^{l+m}}{\ud x^{l+m}} \left(x^2-1\right)^l.
\end{equation}

\subsection{The Radial Equation: Absence of Superradiance}
Superradiance is the wave analogue for the Penrose process on Black Hole. If a bosonic or fermionic wave is incident upon a black hole, normally the reflected wave carries less energy than the incident wave. But under certain condition the transmitted wave, absorbed by the black hole carries negative energy into the black hole making the reflexion coefficient for the wave greater than unity. This means that the reflected wave will carry more energy than the incident wave. This phenomenon is called superradiance by Misner~\cite{dewitt} and also analyzed by Zel'dovich and Starobinsky~\cite{zeldovich,starobinsky}. Through this process energy can be extracted from a black hole in expense of its angular momentum. The condition is given by~\cite{wald}

\begin{equation}
 0<\omega<m\Omega_H,
\end{equation}
 where $\Omega_H$ is the angular velocity of the horizon~\cite{dewitt}. 
Chandrasekhar~\cite{chandra} has analyzed the phenomenon for Kerr geometry in characteristic details and  showed that this occurs only for incident waves of integral spins, i.e., for scalar, electromagnetic waves and gravitational cases. He showed its absence for the fermionic waves, i.e. Dirac wave or neutrino waves. Basak and Majumdar discussed this phenomenon for acoustic analogue of Kerr Black Hole~\cite{bm1,bm2}. 
Clearly for a non-rotating black hole superradiance should be absent as we demonstrate precisely in this case of dyadosphere geometry.

The radial equation is given by
\begin{equation}
 \Delta \frac{\ud}{\ud r}\left( \Delta \frac{\ud R}{\ud r} \right) + \left( \omega^2 r^4 - l(l+1)\Delta\right)R=0.
\end{equation}

Let us introduce the familiar \rs coordinate (the tortoise coordinate) defined by
\[
\frac{\ud\rs}{\ud r} = \frac{r^2}{\Delta},
\]
thus giving
\[
\Delta \frac{\ud }{\ud r} = r^2 \frac{\ud }{\ud \rs}.
\]

Note that though the variable \rs is defined in the same manner as in Schwarzschild or in Kerr metric, in this case the variable is non-integrable. Still the basic purpose is satisfied, the coordinate spans over the real line and pushes the horizon to minus infinity.

The introduction of another function $u(r) = rR$ reduces the radial equation in a more   familiar form
\begin{equation}
\frac{\ud^2 u}{\ud {\rs} ^2} - \left[ \frac{4}{5}\sigma Q^4 \frac{\Delta}{r^{10}} - \frac{2 \Delta^2}{r^6} - \frac{ 2\Delta M}{r^5} + (2 + l(l+1))\frac{\Delta}{r^4} - \omega^2\right] u = 0.
\end{equation}

Thus a potential barrier remains where

\begin{equation}
V(r) = \frac{4}{5}\sigma Q^4 \frac{\Delta}{r^{10}} - \frac{2 \Delta^2}{r^6} - \frac{ 2\Delta M}{r^5} + (2 + l(l+1))\frac{\Delta}{r^4} - \omega^2.
\end{equation}

At horizon ($\Delta \rightarrow 0$), the radial equation becomes
\begin{equation}
 \frac{\ud^2 u_H}{\ud {\rs}^2} + \omega^2 u_H =0,
\end{equation}
with $V(r) = -\omega^2$.

Now asymptotically, $r \rightarrow \infty$. The equation has the same form as in the previous case
\begin{equation}
 \frac{\ud^2 u_\infty}{\ud {\rs}^2} + \omega^2 u_\infty =0.
\end{equation}

Thus $u_H = u_\infty$, where $u_H$ is the radial solution at horizon and $u_\infty$ is the solution at $\infty$. This equality shows that for a dyadosphere metric there is \textit{no phenomenon of superradiance for an incident massless scalar field}. This is expected as dyadosphere geometry is nonrotating.


\section{Analysis of Dyadosphere metric using Lagrangian Method}\label{lagrangian}
The dyadosphere geometry evidently allows two timelike Killing vectors $\partial_t$ and $\partial_\phi$. Thus there must be two constants of motion.  Note that the geodesic in space time can be derived from the usual Lagrangian~\cite{chandra} 
\begin{equation}
 2 \mathcal{L} = g_{ij} \frac{\ud x^i}{\ud \tau}\frac{\ud x^j}{\ud \tau}.
\end{equation}

Here $\tau$ is some affine parameter along the geodesic. For time-like geodesic $\tau$ may be identified with the proper time $s$ of the particle describing the geodesic.

For the Dyadosphere metric the Lagrangian is thus given by
\begin{equation}
 2 \mathcal{L} = f(r) \dot{t}^2 - \frac{1}{f(r)} \dot{r} - r^2 \dot{\theta}^2 - r^2, \sin^2\theta \dot{\phi}^2
\end{equation}
with the dot meaning the differention with respect to $\tau$.

The canonical momenta are given by 
\[
 p_j = \tfrac{\mathcal{L}}{\dot{q_j}}.
\]
The momenta in this case are given by~\cite{chandra}
\begin{equation}\label{conjmom}
 \begin{array}{l}
 p_t = \tfrac{\mathcal{L}}{\dot{t}} = f(r) \dot{t},\\
\\
p_r = - \tfrac{\mathcal{L}}{\dot{r}} = + \frac{1}{f(r)} \dot{r},\\
\\
p_\theta = r^2 \dot{\theta},\\
\\
p_\phi= - \tfrac{\mathcal{L}}{\dot{\phi}} = r^2 \sin^2 \theta \dot{\phi}.
\end{array}
\end{equation}

The Hamiltonian is given by
\begin{equation}
 \mathcal{H} = p_t \dot{t} - \left(p_r \dot{r} + p_\theta \dot{\theta}+ p_\phi \dot{\phi} \right) - \mathcal{L}.
\end{equation}

Putting the values of the conjugate momenta from Eq.~\eqref{conjmom} we find that
\[
 \mathcal{H} = \mathcal{L},
\]
which means the problem has no potential energy term.

Rescaling of the affine parameter $\tau$ allows us to set $2\mathcal{L}=1$ for time-like geodesics. For null geodesics $\mathcal{L}=0$. 

As the metric allows two timelike Killing vectors $\partial_t$ and $\partial_\phi$ so we shall have two constants of integration, namely the conjugate momenta corresponding to the coordinates $t$ and $\phi$. Thus
\begin{equation}
 \begin{array}{l}
 \bgfrac{\ud p_t}{\ud \tau}= \tfrac{\mathcal{L}}{t}=0,\\
\\
\bgfrac{\ud p_\phi}{\ud \tau} = -\tfrac{\mathcal{L}}{\phi} =0.
\end{array}
\end{equation}

Thus we find the first constant of motion
\begin{equation}
 p_t = f(r) \frac{\ud t}{\ud \tau} = \text{constant} = E \text{(say)}.
\end{equation} 
 The metric is asymptotically flat. So the quantity $E$ can be identified to the energy at infinity.

The second constant of motion is given by
\begin{equation}
 p_\phi = r^2 \sin^2 \theta \frac{\ud \phi}{\ud \tau} = \text{constant}.
\end{equation}

From the equation of motion we obtain
\begin{equation}
\frac{\ud p_\theta}{\ud \theta} = \frac{\ud}{\ud \tau}(r^2 \dot{\theta}) = - \tfrac{\mathcal{L}}{\theta} = (r^2 \sin\theta \cos\theta)\left( \frac{\ud \phi}{\ud \tau} \right)^2.
\end{equation} 
 
Note that if we assign  $\theta=\pi/2$ when $\dot{\theta}=0$ then $\ddot{\theta}=0$. This means that the geodesic is described in an invariant plane which may be distinguished by $\theta=\pi/2$. Thus we obtain
\begin{equation}
p_\phi= r^2\frac{\ud \phi}{\ud \tau} = \text{constant} = L \ \text{(say)}.
 \end{equation} 
    
Here $L$ denotes the angular momentum about an axis normal to the invariant plane.

Putting all these constants we obtain
\begin{equation}
 \frac{E^2}{f(r)} - \frac{\dot{r}^2}{f(r)} - \frac{L^2}{r^2} = 2 \mathcal{L} = +1 \ \text{or} \ 0.
\end{equation} 
depending whether it is a timelike or null geodesic.

We find
\begin{equation}
 \dot{r}^2 + \left( \frac{L^2}{r^2} + 2 \mathcal{L} \right) f(r) = E^2.
\end{equation} 
Thus we can identify the following quantity at the left hand side of the equation as  effectively to a potential energy term~\cite[p102]{chandra}
\begin{equation}
 V^2_{\text{eff}} = \left( \frac{L^2}{r^2} + 2 \mathcal{L} \right) f(r).
\end{equation}

\subsection{Radial timelike Geodesic, $\mathcal{L}=+1$, $L=0$}
\begin{figure}[htbp]
\begin{center}
\setlength{\unitlength}{0.240900pt}
\ifx\plotpoint\undefined\newsavebox{\plotpoint}\fi
\sbox{\plotpoint}{\rule[-0.200pt]{0.400pt}{0.400pt}}%
\begin{picture}(1500,900)(0,0)
\sbox{\plotpoint}{\rule[-0.200pt]{0.400pt}{0.400pt}}%
\put(140.0,82.0){\rule[-0.200pt]{4.818pt}{0.400pt}}
\put(120,82){\makebox(0,0)[r]{ 0}}
\put(1419.0,82.0){\rule[-0.200pt]{4.818pt}{0.400pt}}
\put(140.0,193.0){\rule[-0.200pt]{4.818pt}{0.400pt}}
\put(120,193){\makebox(0,0)[r]{ 0.1}}
\put(1419.0,193.0){\rule[-0.200pt]{4.818pt}{0.400pt}}
\put(140.0,304.0){\rule[-0.200pt]{4.818pt}{0.400pt}}
\put(120,304){\makebox(0,0)[r]{ 0.2}}
\put(1419.0,304.0){\rule[-0.200pt]{4.818pt}{0.400pt}}
\put(140.0,415.0){\rule[-0.200pt]{4.818pt}{0.400pt}}
\put(120,415){\makebox(0,0)[r]{ 0.3}}
\put(1419.0,415.0){\rule[-0.200pt]{4.818pt}{0.400pt}}
\put(140.0,527.0){\rule[-0.200pt]{4.818pt}{0.400pt}}
\put(120,527){\makebox(0,0)[r]{ 0.4}}
\put(1419.0,527.0){\rule[-0.200pt]{4.818pt}{0.400pt}}
\put(140.0,638.0){\rule[-0.200pt]{4.818pt}{0.400pt}}
\put(120,638){\makebox(0,0)[r]{ 0.5}}
\put(1419.0,638.0){\rule[-0.200pt]{4.818pt}{0.400pt}}
\put(140.0,749.0){\rule[-0.200pt]{4.818pt}{0.400pt}}
\put(120,749){\makebox(0,0)[r]{ 0.6}}
\put(1419.0,749.0){\rule[-0.200pt]{4.818pt}{0.400pt}}
\put(140.0,860.0){\rule[-0.200pt]{4.818pt}{0.400pt}}
\put(120,860){\makebox(0,0)[r]{ 0.7}}
\put(1419.0,860.0){\rule[-0.200pt]{4.818pt}{0.400pt}}
\put(140.0,82.0){\rule[-0.200pt]{0.400pt}{4.818pt}}
\put(140,41){\makebox(0,0){ 0}}
\put(140.0,840.0){\rule[-0.200pt]{0.400pt}{4.818pt}}
\put(400.0,82.0){\rule[-0.200pt]{0.400pt}{4.818pt}}
\put(400,41){\makebox(0,0){ 1}}
\put(400.0,840.0){\rule[-0.200pt]{0.400pt}{4.818pt}}
\put(660.0,82.0){\rule[-0.200pt]{0.400pt}{4.818pt}}
\put(660,41){\makebox(0,0){ 2}}
\put(660.0,840.0){\rule[-0.200pt]{0.400pt}{4.818pt}}
\put(919.0,82.0){\rule[-0.200pt]{0.400pt}{4.818pt}}
\put(919,41){\makebox(0,0){ 3}}
\put(919.0,840.0){\rule[-0.200pt]{0.400pt}{4.818pt}}
\put(1179.0,82.0){\rule[-0.200pt]{0.400pt}{4.818pt}}
\put(1179,41){\makebox(0,0){ 4}}
\put(1179.0,840.0){\rule[-0.200pt]{0.400pt}{4.818pt}}
\put(1439.0,82.0){\rule[-0.200pt]{0.400pt}{4.818pt}}
\put(1439,41){\makebox(0,0){ 5}}
\put(1439.0,840.0){\rule[-0.200pt]{0.400pt}{4.818pt}}
\put(140.0,82.0){\rule[-0.200pt]{312.929pt}{0.400pt}}
\put(1439.0,82.0){\rule[-0.200pt]{0.400pt}{187.420pt}}
\put(140.0,860.0){\rule[-0.200pt]{312.929pt}{0.400pt}}
\put(140.0,82.0){\rule[-0.200pt]{0.400pt}{187.420pt}}
\put(875,107){\usebox{\plotpoint}}
\multiput(875.58,107.00)(0.493,4.739){23}{\rule{0.119pt}{3.792pt}}
\multiput(874.17,107.00)(13.000,112.129){2}{\rule{0.400pt}{1.896pt}}
\multiput(888.58,227.00)(0.493,2.241){23}{\rule{0.119pt}{1.854pt}}
\multiput(887.17,227.00)(13.000,53.152){2}{\rule{0.400pt}{0.927pt}}
\multiput(901.58,284.00)(0.493,1.646){23}{\rule{0.119pt}{1.392pt}}
\multiput(900.17,284.00)(13.000,39.110){2}{\rule{0.400pt}{0.696pt}}
\multiput(914.58,326.00)(0.493,1.408){23}{\rule{0.119pt}{1.208pt}}
\multiput(913.17,326.00)(13.000,33.493){2}{\rule{0.400pt}{0.604pt}}
\multiput(927.58,362.00)(0.493,1.171){23}{\rule{0.119pt}{1.023pt}}
\multiput(926.17,362.00)(13.000,27.877){2}{\rule{0.400pt}{0.512pt}}
\multiput(940.58,392.00)(0.494,0.974){25}{\rule{0.119pt}{0.871pt}}
\multiput(939.17,392.00)(14.000,25.191){2}{\rule{0.400pt}{0.436pt}}
\multiput(954.58,419.00)(0.493,0.933){23}{\rule{0.119pt}{0.838pt}}
\multiput(953.17,419.00)(13.000,22.260){2}{\rule{0.400pt}{0.419pt}}
\multiput(967.58,443.00)(0.493,0.853){23}{\rule{0.119pt}{0.777pt}}
\multiput(966.17,443.00)(13.000,20.387){2}{\rule{0.400pt}{0.388pt}}
\multiput(980.58,465.00)(0.493,0.774){23}{\rule{0.119pt}{0.715pt}}
\multiput(979.17,465.00)(13.000,18.515){2}{\rule{0.400pt}{0.358pt}}
\multiput(993.58,485.00)(0.493,0.734){23}{\rule{0.119pt}{0.685pt}}
\multiput(992.17,485.00)(13.000,17.579){2}{\rule{0.400pt}{0.342pt}}
\multiput(1006.58,504.00)(0.493,0.655){23}{\rule{0.119pt}{0.623pt}}
\multiput(1005.17,504.00)(13.000,15.707){2}{\rule{0.400pt}{0.312pt}}
\multiput(1019.58,521.00)(0.493,0.655){23}{\rule{0.119pt}{0.623pt}}
\multiput(1018.17,521.00)(13.000,15.707){2}{\rule{0.400pt}{0.312pt}}
\multiput(1032.58,538.00)(0.493,0.576){23}{\rule{0.119pt}{0.562pt}}
\multiput(1031.17,538.00)(13.000,13.834){2}{\rule{0.400pt}{0.281pt}}
\multiput(1045.58,553.00)(0.493,0.536){23}{\rule{0.119pt}{0.531pt}}
\multiput(1044.17,553.00)(13.000,12.898){2}{\rule{0.400pt}{0.265pt}}
\multiput(1058.00,567.58)(0.497,0.494){25}{\rule{0.500pt}{0.119pt}}
\multiput(1058.00,566.17)(12.962,14.000){2}{\rule{0.250pt}{0.400pt}}
\multiput(1072.00,581.58)(0.497,0.493){23}{\rule{0.500pt}{0.119pt}}
\multiput(1072.00,580.17)(11.962,13.000){2}{\rule{0.250pt}{0.400pt}}
\multiput(1085.00,594.58)(0.539,0.492){21}{\rule{0.533pt}{0.119pt}}
\multiput(1085.00,593.17)(11.893,12.000){2}{\rule{0.267pt}{0.400pt}}
\multiput(1098.00,606.58)(0.539,0.492){21}{\rule{0.533pt}{0.119pt}}
\multiput(1098.00,605.17)(11.893,12.000){2}{\rule{0.267pt}{0.400pt}}
\multiput(1111.00,618.58)(0.590,0.492){19}{\rule{0.573pt}{0.118pt}}
\multiput(1111.00,617.17)(11.811,11.000){2}{\rule{0.286pt}{0.400pt}}
\multiput(1124.00,629.58)(0.590,0.492){19}{\rule{0.573pt}{0.118pt}}
\multiput(1124.00,628.17)(11.811,11.000){2}{\rule{0.286pt}{0.400pt}}
\multiput(1137.00,640.58)(0.652,0.491){17}{\rule{0.620pt}{0.118pt}}
\multiput(1137.00,639.17)(11.713,10.000){2}{\rule{0.310pt}{0.400pt}}
\multiput(1150.00,650.58)(0.652,0.491){17}{\rule{0.620pt}{0.118pt}}
\multiput(1150.00,649.17)(11.713,10.000){2}{\rule{0.310pt}{0.400pt}}
\multiput(1163.00,660.59)(0.786,0.489){15}{\rule{0.722pt}{0.118pt}}
\multiput(1163.00,659.17)(12.501,9.000){2}{\rule{0.361pt}{0.400pt}}
\multiput(1177.00,669.59)(0.728,0.489){15}{\rule{0.678pt}{0.118pt}}
\multiput(1177.00,668.17)(11.593,9.000){2}{\rule{0.339pt}{0.400pt}}
\multiput(1190.00,678.59)(0.728,0.489){15}{\rule{0.678pt}{0.118pt}}
\multiput(1190.00,677.17)(11.593,9.000){2}{\rule{0.339pt}{0.400pt}}
\multiput(1203.00,687.59)(0.824,0.488){13}{\rule{0.750pt}{0.117pt}}
\multiput(1203.00,686.17)(11.443,8.000){2}{\rule{0.375pt}{0.400pt}}
\multiput(1216.00,695.59)(0.824,0.488){13}{\rule{0.750pt}{0.117pt}}
\multiput(1216.00,694.17)(11.443,8.000){2}{\rule{0.375pt}{0.400pt}}
\multiput(1229.00,703.59)(0.824,0.488){13}{\rule{0.750pt}{0.117pt}}
\multiput(1229.00,702.17)(11.443,8.000){2}{\rule{0.375pt}{0.400pt}}
\multiput(1242.00,711.59)(0.950,0.485){11}{\rule{0.843pt}{0.117pt}}
\multiput(1242.00,710.17)(11.251,7.000){2}{\rule{0.421pt}{0.400pt}}
\multiput(1255.00,718.59)(0.824,0.488){13}{\rule{0.750pt}{0.117pt}}
\multiput(1255.00,717.17)(11.443,8.000){2}{\rule{0.375pt}{0.400pt}}
\multiput(1268.00,726.59)(1.026,0.485){11}{\rule{0.900pt}{0.117pt}}
\multiput(1268.00,725.17)(12.132,7.000){2}{\rule{0.450pt}{0.400pt}}
\multiput(1282.00,733.59)(1.123,0.482){9}{\rule{0.967pt}{0.116pt}}
\multiput(1282.00,732.17)(10.994,6.000){2}{\rule{0.483pt}{0.400pt}}
\multiput(1295.00,739.59)(0.950,0.485){11}{\rule{0.843pt}{0.117pt}}
\multiput(1295.00,738.17)(11.251,7.000){2}{\rule{0.421pt}{0.400pt}}
\multiput(1308.00,746.59)(1.123,0.482){9}{\rule{0.967pt}{0.116pt}}
\multiput(1308.00,745.17)(10.994,6.000){2}{\rule{0.483pt}{0.400pt}}
\multiput(1321.00,752.59)(1.123,0.482){9}{\rule{0.967pt}{0.116pt}}
\multiput(1321.00,751.17)(10.994,6.000){2}{\rule{0.483pt}{0.400pt}}
\multiput(1334.00,758.59)(1.123,0.482){9}{\rule{0.967pt}{0.116pt}}
\multiput(1334.00,757.17)(10.994,6.000){2}{\rule{0.483pt}{0.400pt}}
\multiput(1347.00,764.59)(1.123,0.482){9}{\rule{0.967pt}{0.116pt}}
\multiput(1347.00,763.17)(10.994,6.000){2}{\rule{0.483pt}{0.400pt}}
\multiput(1360.00,770.59)(1.123,0.482){9}{\rule{0.967pt}{0.116pt}}
\multiput(1360.00,769.17)(10.994,6.000){2}{\rule{0.483pt}{0.400pt}}
\multiput(1373.00,776.59)(1.489,0.477){7}{\rule{1.220pt}{0.115pt}}
\multiput(1373.00,775.17)(11.468,5.000){2}{\rule{0.610pt}{0.400pt}}
\multiput(1387.00,781.59)(1.123,0.482){9}{\rule{0.967pt}{0.116pt}}
\multiput(1387.00,780.17)(10.994,6.000){2}{\rule{0.483pt}{0.400pt}}
\multiput(1400.00,787.59)(1.378,0.477){7}{\rule{1.140pt}{0.115pt}}
\multiput(1400.00,786.17)(10.634,5.000){2}{\rule{0.570pt}{0.400pt}}
\multiput(1413.00,792.59)(1.378,0.477){7}{\rule{1.140pt}{0.115pt}}
\multiput(1413.00,791.17)(10.634,5.000){2}{\rule{0.570pt}{0.400pt}}
\multiput(1426.00,797.59)(1.378,0.477){7}{\rule{1.140pt}{0.115pt}}
\multiput(1426.00,796.17)(10.634,5.000){2}{\rule{0.570pt}{0.400pt}}
\put(140.0,82.0){\rule[-0.200pt]{312.929pt}{0.400pt}}
\put(1439.0,82.0){\rule[-0.200pt]{0.400pt}{187.420pt}}
\put(140.0,860.0){\rule[-0.200pt]{312.929pt}{0.400pt}}
\put(140.0,82.0){\rule[-0.200pt]{0.400pt}{187.420pt}}
\end{picture}
\end{center}
\caption{$V_{\text{eff}}$ with $M=1.5$, $Q=0.7$ and $\sigma = 0.5$}\label{figgeotime}
\end{figure}
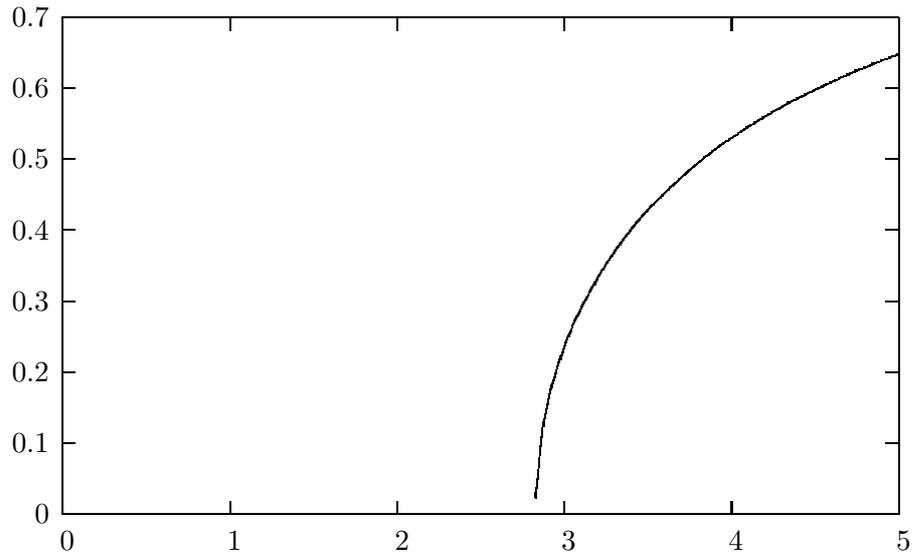

The effective potential for radial geodesic (zero angular momentum) is given by

\begin{equation}
V^2_{\text{eff}} = f(r) = 1 - \frac{2M}{r} + \frac{Q^2}{r^2} - \frac{\sigma Q^4}{5 r^6},
\end{equation}
 which is equal to the original dyadosphere metric. The effective potential is plotted (Fig.~\ref{figgeotime}) here for a black hole of  $M=1.5$, $Q=0.04$, and $\sigma =0.5$.

\subsection{Null geodesic}
For null geodesic, $ \mathcal{L}=0$ and the effective potential is given by
\begin{equation}
V^2_{\text{eff}}=\frac{L^2}{r^2×} f(r).
\end{equation}
 
\begin{figure}[!ht]
\begin{center}
\input{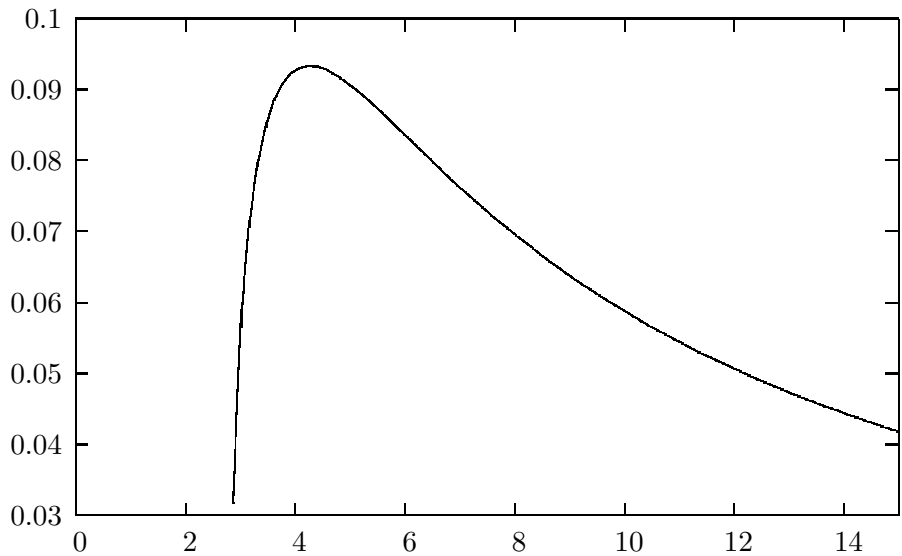}
\end{center}
\caption{$V_{\text{eff}}$ for $L=0.7$, $M=1.5$, $Q=0.7$, and $\sigma = 0.5$}\label{figgeonull}
\end{figure}

 Thus the effective potential for a photon with zero angular momentum vanishes. 
The effective potential is plotted (Fig.~\ref{figgeonull}) for a black hole of $M=1.5$, $Q=0.7$, $\sigma=5$ and $L=0.7$.

\section{Discussion}
In this paper we discussed motion of test particle in the dyadosphere metric using Hamilton-Jacobi techniques. Some representative curves for effective potential is plotted. (All the curves are drawn by \texttt{gnuplot} plotting program). Several plots can be generated by changing the values of the parameters to see the nature of the function in different cases. We also discussed geodesics through Lagrangian method.  The wave equation for a massless particle also is studied to show that the phenomenon of superradiance is absent in this case.

\section*{Acknowledgment} BR likes to thank The Inter-University Centre for Astronomy and Astrophysics (IUCAA), Pune, India for excellent facilities extended to him during a visit under the Associateship programme.

\end{document}